\documentclass{ISPIV2025_Template_Paper_Class}

\usepackage{placeins}
\begin{document}
\title{FED-PV: A Large-Scale Synthetic Frame/Event Dataset for Particle-Based Velocimetry}
\author{Fan Wu, Xiang Feng, Aoyu Zhang, Yong Lee$^{*}$}
\affiliation{
Hubei Provincial Engineering Research Center of Robotics \& Intelligent Manufacturing, School of Mechanical and Electronic Engineering, Wuhan University of Technology, Wuhan 430070, China\\
		\vspace{7pt}
		$^*$ yonglee@whut.edu.cn
		}
\maketitle

\begin{abstract}
Particle-based velocimetry (PV) is a widely used technique for non-invasive flow field measurements in fluid mechanics. Existing PV measurements typically rely on a single type of particle recording. With advancements in deep learning and information fusion, incorporating multiple different particle recordings presents a promising avenue for next-generation PV measurement techniques. However, we argue that the lack of cross-modal datasets---combining frame-based recordings and event-based recordings---represents a significant bottleneck in the development of fusion measurement algorithms. To address this critical gap, we developed a dual-modal data generator FED-PV to synthesize frame-based images and event-based recordings of moving particles, resulting in a $350GB$ dataset generated using our approach. This generator and dataset will facilitate advancements in novel PV algorithms.


\end{abstract}

\section{Introduction}

Particle-based velocimetry (PV) is a non-invasive flow field measurement technique widely used in experimental fluid dynamics, including particle image velocimetry~(PIV)~\cite{raffel2018particleguide,lee2017piv,ai2025rethinking}, particle tracking velocimetry~(PTV)~\cite{ohmi2000particle}, event-based imaging velocimetry~(EBIV)~\cite{willert2022event}, particle event velocimetry~(PEV)~\cite{alsattam2024kf}.
However, as the demands for higher performance, the limitations of single-modal PV methods have become apparent, making it challenging to improve the accuracy of measurements further~\cite{kahler2012resolutionPIVlimit,lee2024surrogate,lee2021diffeomorphic,ai2025projection}. In particular, research on integrating multiple data sources, especially combining event-based data with particle image data, remains in its infancy and has yet to fully realize its potential~\cite{wan2023rpeflow}. Thus, constructing a dual-modal particle velocity dataset not only fills the critical gap but also provides training/testing benchmarks for advancing PV technology in the coming measurement studies.

Constructing multimodal datasets (e.g., RGB and event data) is a common practice in the computer vision community. The datasets can be derived from real-world recordings~\cite{zhu2018multivehicle,gehrig2021dsec} or generated using various simulation tools~\cite{mueggler2017eventsimulator,rebecq2018esim}. However, these readily available datasets are rooted in real-world scenarios and cannot be directly applied to the PV area due to the domain gap. Due to the unknown latent flow fields, it is also challenging to obtain a large amount of dual-modal data with ground truth using real-world setups.

To overcome these limitations, we developed a particle frame/event generator and constructed a dual-modal particle flow dataset with ground truth.  Specifically, the flow fields are derived from a publicly available dataset, encompassing diverse types of flows as described in \cite{cai2019dense}. Then, our simulator disperses particles within a given flow field, generating a high-frame-rate particle flow video~\cite{raffel2018particleguide}. The event data can be extracted from the video, while simultaneously, images at fixed time intervals are selected for PIV analysis.
The primary contributions are as follows:
\begin{itemize}
\item A particle frame/event generator. We provide an integrated tool that synthesizes particle image sequences and event stream data based on a specified velocity field. The generator is highly customizable, allowing users to adjust key parameters such as particle density and event triggering thresholds to meet diverse data generation requirements.
\item A dual-modal dataset. The dataset--- integrating particle images, events, and ground truth flow fields--- is released.  The dataset includes nine distinct flow field types and has a total size of $350GB$, making it one dual-modal dataset available for advanced particle-based velocimetry.
\end{itemize}

\section{Methods for Dataset Generation}
\subsection{Overview of Our Dataset}

\begin{figure}[!hptb]
    \centering
    \includegraphics[height=.3\textwidth]{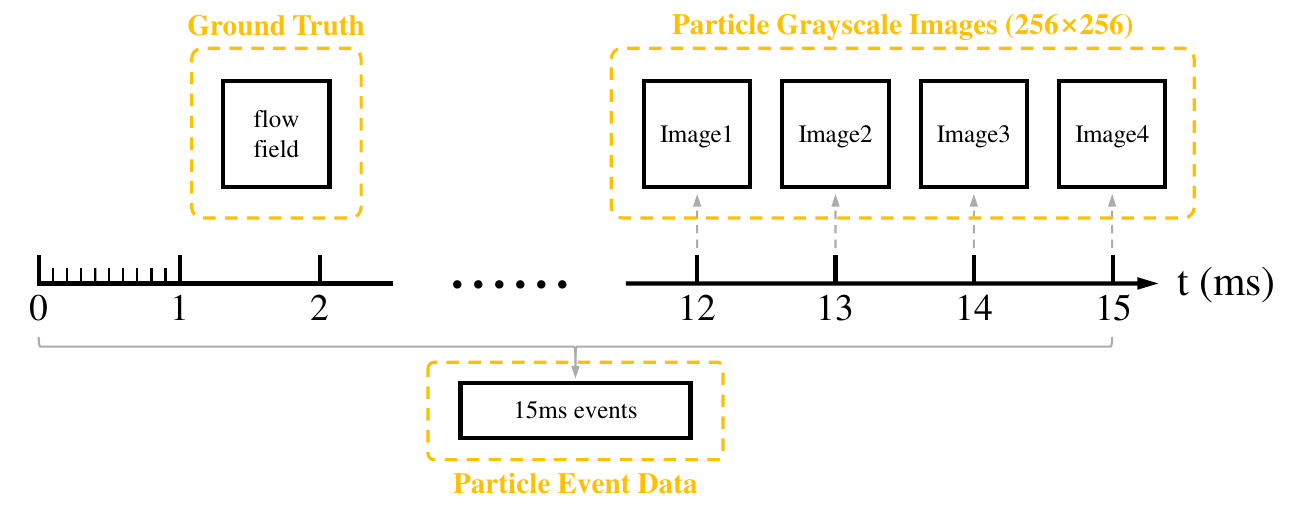}
    \caption{Timing diagram for FED-PV.}
    \label{fig:Timeline}
\end{figure}

This section details the structure of the released dataset and the computational methods used for data generation, with particular emphasis on the technical implementation.
Our dataset is based on the widely used dataset in the field of Particle Image Velocimetry (PIV) \cite{cai2019dense}, which includes nine distinct flow field types with diverse dynamics. Building upon the work of Cai et al., we redefined the particle velocity units to meet the requirements for event data recording. For each flow field, event data was recorded over a duration of $15$ ms, with the final $4$ grayscale images preserved at $1$ ms intervals (each image represents a snapshot of the corresponding moment). The timeline in Figure\,\ref{fig:Timeline} provides a detailed explanation of the temporal sequence of particle image generation and event data acquisition.

\subsection{The Image Generation Method}
We utilized the CFD flow motion patterns from a publicly available dataset \cite{cai2019dense} as the ground truth benchmark. The static characteristics of particles are modeled using a two-dimensional(2D) Gaussian function, where the intensity distribution of each particle is represented as:

\begin{equation}
I\left( {x,y}\right)  = {I}_{0}\exp \left\lbrack  \frac{-{\left( x - {x}_{0}\right) }^{2} - {\left( y - {y}_{0}\right) }^{2}}{\left( {1/8}\right) {d}_{\mathrm{p}}^{2}}\right\rbrack \
\end{equation}

\noindent where $(x_0,y_0)$ represents the center position of particle, $I_0$ denotes the peak intensity at the Gaussian center, and $d_p$ is the diameter of particle. The parameters are randomly sampled from predefined normal distributions during initialization, with a particle spatial density set at $0.06$ particles per pixel ($ppp$).

To enhance image quality over long time spans, particularly at the domain boundaries, the computational domain is expanded from the original $256\times256$ flow field to a $288\times288$ spatial region using the smoothing algorithm \cite{Lee_2017,GARCIA20101167}. 

The generation of particle grayscale images follows the Particle Image Generator (PIG) standard \cite{raffel2018particleguide}, producing image sequences at fixed $1ms$ intervals using an iterative process. The final dataset retains the last four frames of the PIV image sequence. Within the iterative process, computations are performed using a minimal time step of $0.02ms$, during which particle motion is approximated as uniform linear movement. The velocity of each particle is equal to the local flow velocity at its position, which is obtained via bilinear interpolation. 

\begin{figure}[!hptb]
    \centering
    \includegraphics[height=.3\textwidth]{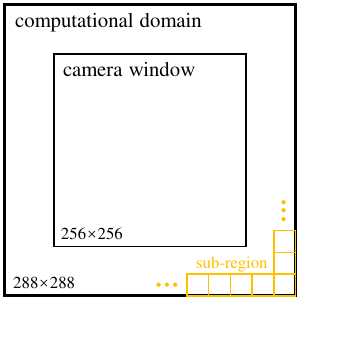}
    \caption{Definition of the computational domain of the flow field.}
    \label{fig:ComputationalDomain}
\end{figure}

To maintain the continuity of particle distribution within the camera window over extended time spans, a dynamic management mechanism for edge sub-regions is introduced at the computational domain boundaries. Figure\,\ref{fig:ComputationalDomain} provides an overview of the computational domain. When the average brightness of an edge sub-region falls below a predefined threshold, new particles are introduced to simulate a continuous influx at a constant density. Meanwhile, to optimize computational efficiency, particles that exceed the computational domain are automatically removed.

\subsection{The Event Generation Method}
\begin{figure}[!hptb]
        \centering
    	 \includegraphics[height=.5\textwidth]{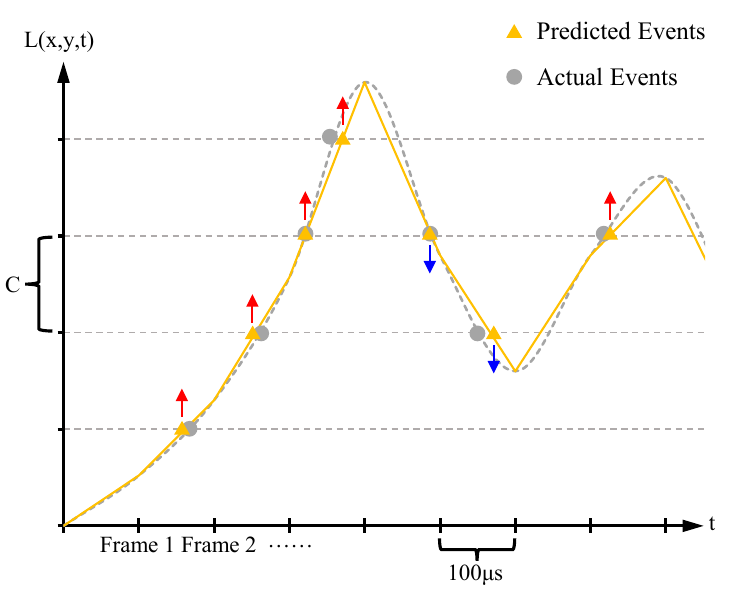}
	 \caption{The Principle of Event Data Generation for the Dataset}
        \label{fig:Eventline}
\end{figure}

Event data is one of the core components of the entire dataset. We extract the brightness information of each pixel from a high-frame-rate image stream to simulate the operation of an event camera~\cite{mueggler2017eventsimulator,rebecq2018esim}. The event generation process is illustrated in the figure (Fig.~\ref{fig:Eventline}). The $x$-axis represents the reading time of each frame of the image, while the $y$-axis indicates the brightness change of a specific pixel. The blue dashed line represents the actual brightness variation, and the yellow solid line shows the piecewise linear fit of the brightness for each frame of the image. $C$ is the threshold, and an event is generated when the brightness change satisfies the following condition:


\begin{equation}
\left|\log \left(x, y,t_{1}\right)-\log \left(x, y, t_{2}\right)\right| \geqslant C
\end{equation}
where $x,y$ represents the pixel position, $t$ represents time, and $C$ is the threshold. A positive event is generated when the brightness change exceeds the threshold, while a negative event is generated when the brightness change is below the threshold.

Typically, the threshold $C$ for event cameras is set within the range of $10\%$ to $50\%$ of the brightness change. A lower threshold results in a high density of events, but this also increases noise interference. On the other hand, a higher threshold effectively suppresses noise, but it also leads to a decrease in event count, which may hinder the detection of subtle changes. Since our dataset does not include noise, we have empirically selected a balanced threshold of $25\%$, where events are triggered when the brightness change exceeds this value. For those interested in exploring the optimal threshold selection, readers can test it themselves using our open-source code.


To ensure that high-frame-rate image data can accurately capture the motion trajectory of each particle, our dataset adopts an inter-frame interval of $100\mu s$. Given that particle velocities in the dataset typically range from $3$ to $9 pixels/ms$, the displacement between two consecutive frames does not exceed $1 pixel$. This enables precise recording of the brightness variations across individual pixel locations traversed by the particles, significantly improving the accuracy of linear interpolation and thereby ensuring high precision in event data generation.


\subsection{Details of a released Dataset}
The dataset is categorized into $9$ subsets based on the dynamic characteristics of flow fields, each corresponding to a distinct type of velocity field (the subset names are strictly aligned with the respective velocity field types). Within subsets, multiple scenarios are provided, each characterized by different flow field parameters and initial conditions. A scenario comprises four types of data: (1) Ground-truth velocity fields (stored in $.flo$ format with units in pixel/s); (2) PIV images, comprising four sequential frames; (3) Event stream data (stored in $.h5$ format). All images are single-channel grayscale images ($.png$ format) with $256\times256$ pixel resolution. A comprehensive overview of the quantity and size of the dataset is presented in Table\,\ref{dataset_info}. Compared to the dataset by Cai et al., our dataset contains a greater number of particle grayscale images and includes a substantial amount of event-based data.

\begin{table}[hptb]
\centering
\caption{Dataset scale information}
\label{dataset_info}
\begin{tblr}{
  hline{1,11} = {-}{0.08em},
  hline{2} = {-}{},
}
Flow type                & Quantity & {Average velocity\\$(pixel/ms)$} & {Average number \\of events}       & Size   \\
backstep                 & $3600$   & $1.8-4.4$                        & $6.13\times10^6$\textsuperscript{} & $94G$  \\
cylinder                 & $2050$   & $0.9-1.8$                        & $2.61\times10^6$\textsuperscript{} & $27G$  \\
DNS turbulence           & $2000$   & $0.2-2.9$                        & $2.88\times10^6$\textsuperscript{} & $29G$  \\
JHTDB\_channel           & $1900$   & $0.8-6.7$                        & $8.92\times10^6$\textsuperscript{} & $81G$  \\
JHTDB\_channel \_hd      & $600$    & $0.2-1.0$                        & $1.22\times10^6$\textsuperscript{} & $3.9G$ \\
JHTDB\_isotropic1024\_hd & $2000$   & $0.3-5.6$                        & $2.83\times10^6$\textsuperscript{} & $28G$  \\
JHTDB\_mhd1024\_hd       & $800$    & $0.4-3.5$                        & $3.17\times10^6$\textsuperscript{} & $13G$  \\
SQG                      & $1500$   & $1.3-1.8$                        & $3.46\times10^6$\textsuperscript{} & $26G$  \\
uniform                  & $1000$   & $0.1-8.3$                        & $1.00\times10^7$                   & $48G$  
\end{tblr}
\end{table}

\begin{figure}[hptb]
        \centering
    	 \includegraphics[width=\textwidth]{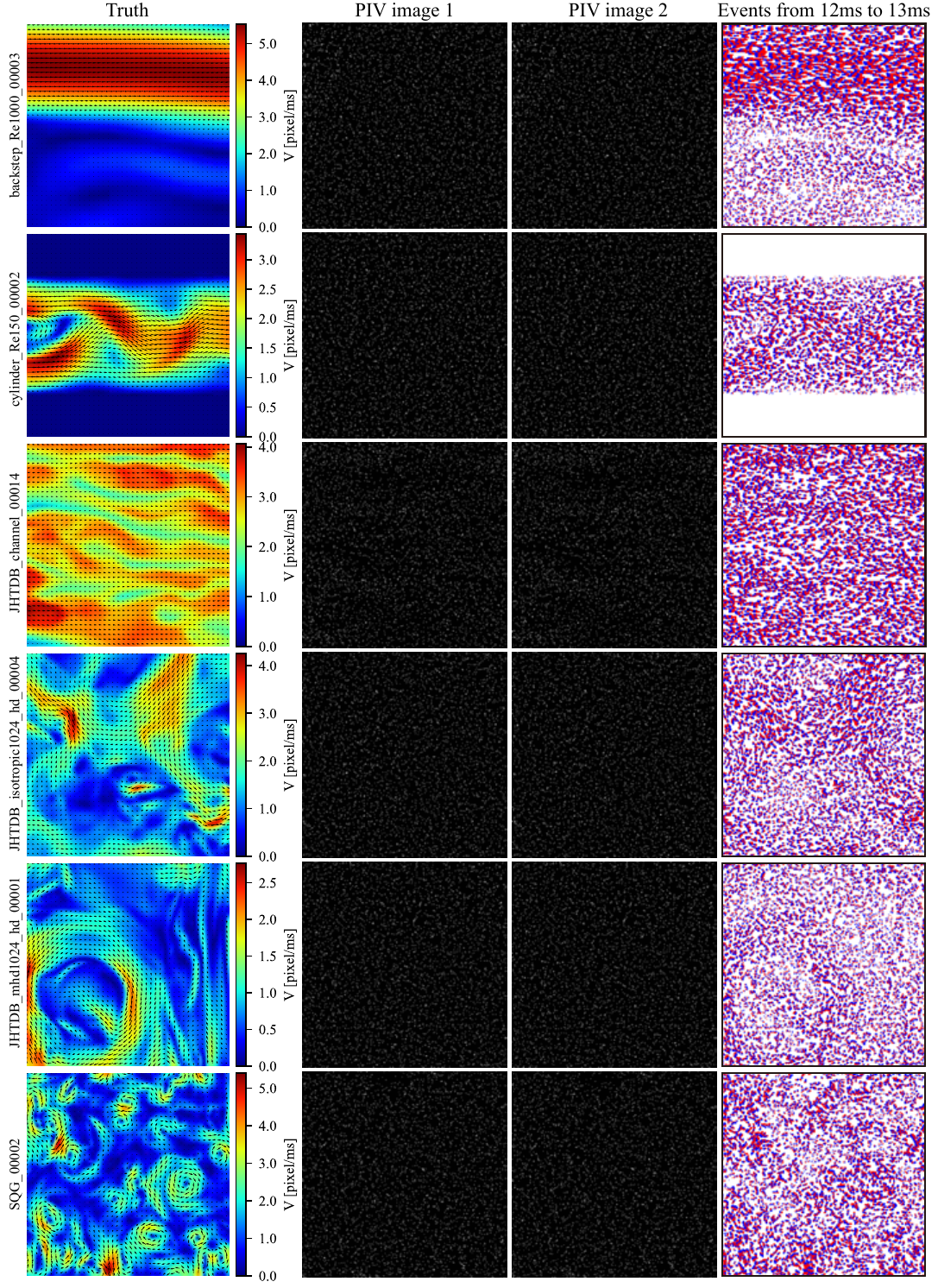}
	 \caption{The visualization of our dataset.}
        \label{fig:ShowDtaset}
\end{figure}

\section{Benchmarking of the Dataset}
In this section, we evaluate the reliability and usability of our dataset using three mainstream PIV algorithms and three event-based algorithms. Based on these methods, we establish a benchmark on our dataset to provide a reference for future research and facilitate its application by other users.


\subsection{PIV Application}
To evaluate the quality of the image data generated in our dataset, we conducted tests using three representative methods: UnLiteFlowNet-PIV\cite{zhang2020unsupervised}, Cross-correlation-PIV\cite{Liberzon2016OpenpivOpenpivPythonUP}  , and RAFT-PIV \cite{teed2020raft}. The images used in all tests were Image1 and Image2 for each flow field (corresponding to the flow fields at 12ms and 13ms, respectively), with an image resolution of $256 \times 256$.

\begin{figure}[hptb]
        \centering
    	 \includegraphics[width=\textwidth]{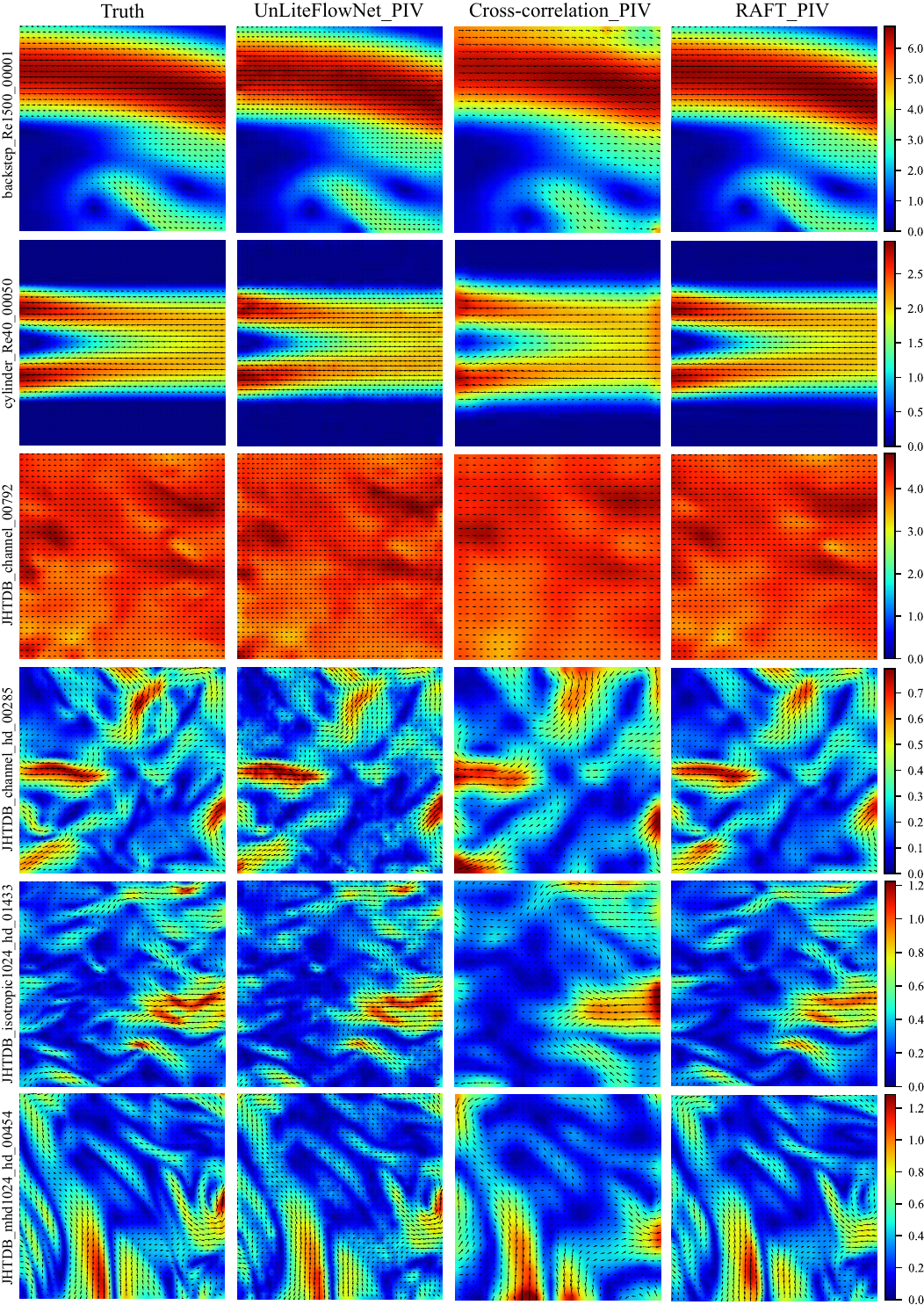}
	 \caption{The visualization of PIV Application.}
        \label{fig:PIV_results}
\end{figure}

Figure~\ref{fig:PIV_results} presents the testing results of the six flow fields using the three aforementioned methods. Overall, all three methods are capable of reasonably reconstructing the underlying flow structures. Specifically, the UnLiteFlowNet-PIV method exhibits noticeable discontinuities in the estimated flow, such as abrupt changes in the bottom-right corner of JHTDB\_channel\_hd\_00285. The Cross-correlation-PIV method, while limited in resolution due to its inherent computational constraints, still captures the general flow patterns present in the dataset. RAFT-PIV, as a well-established baseline, produces consistently high-quality results with superior accuracy. These observations align well with existing knowledge in the PIV research community and further demonstrate the reliability of our dataset for PIV-based flow analysis.

In addition, we adopt three commonly used evaluation metrics in the PIV domain—RMSE, AEE, and AAE—to establish a benchmark for our dataset. Table~\ref{Tab:PIV_assess} presents the statistical characteristics of these metrics across nine types of flow fields in our dataset, evaluated using UnLiteFlowNet-PIV, Cross-correlation-PIV, and RAFT-PIV. For most flow fields, the RMSE values of the three methods fall within the range of $0.1$ to $0.3$. However, in more complex flow scenarios, Cross-correlation-PIV tends to exhibit larger errors. We hope these results can serve as a useful reference for future algorithm development based on our dataset. Meanwhile, the notable room for improvement observed across multiple flow fields indicates that there remains significant potential for advancing high-performance PIV algorithms.


\begin{table}
\centering
\caption{Quantitative comparison of different PIV methods evaluated on our proposed dataset. }
\label{Tab:PIV_assess}
\begin{tblr}{
  colspec = {llccc},
  cell{2}{1} = {r=3}{},
  cell{5}{1} = {r=3}{},
  cell{8}{1} = {r=3}{},
  cell{11}{1} = {r=3}{},
  cell{14}{1} = {r=3}{},
  cell{17}{1} = {r=3}{},
  cell{20}{1} = {r=3}{},
  cell{23}{1} = {r=3}{},
  cell{26}{1} = {r=3}{},
  hline{1,29} = {-}{0.08em},
  hline{2,5,8,11,14,17,20,23,26} = {-}{},
}
Flow type~                & Method                & RMSE  & AEE   & AAE   \\
backstep~                 & UnLiteFlowNet\_PIV    & $0.134$ & $0.058$ & $0.056$ \\
                          & Cross-correlation\_PIV~ & $0.201$ & $0.101$ & $0.033$ \\
                          & RAFT\_PIV             & $0.167$ & $0.107$ & $0.030$ \\
cylinder~                 & UnLiteFlowNet\_PIV    & $0.075$ & $0.058$ & $0.683$ \\
                          & Cross-correlation\_PIV~ & $0.219$ & $0.143$ & $0.593$ \\
                          & RAFT\_PIV             & $0.089$ & $0.065$ & $0.684$ \\
DNS\_turbulence~          & UnLiteFlowNet\_PIV    & $0.206$ & $0.165$ & $0.137$ \\
                          & Cross-correlation\_PIV~ & $1.037$ & $0.807$ & $0.577$ \\
                          & RAFT\_PIV             & $0.218$ & $0.163$ & $0.111$ \\
JHTDB\_channel~           & UnLiteFlowNet\_PIV    & $0.147$ & $0.095$ & $0.018$ \\
                          & Cross-correlation\_PIV~ & $0.305$ & $0.221$ & $0.028$ \\
                          & RAFT\_PIV             & $0.268$ & $0.214$ & $0.041$ \\
JHTDB\_channel\_hd~       & UnLiteFlowNet\_PIV    & $0.089$ & $0.073$ & $0.129$ \\
                          & Cross-correlation\_PIV~ & $0.240$ & $0.189$ & $0.304$ \\
                          & RAFT\_PIV             & $0.122$ & $0.100$ & $0.186$ \\
JHTDB\_isotropic1024\_hd~ & UnLiteFlowNet\_PIV    & $0.157$ & $0.119$ & $0.097$ \\
                          & Cross-correlation\_PIV~ & $0.441$ & $0.331$ & $0.249$ \\
                          & RAFT\_PIV             & $0.336$ & $0.252$ & $0.176$ \\
JHTDB\_mhd1024\_hd~       & UnLiteFlowNet\_PIV    & $0.154$ & $0.118$ & $0.083$ \\
                          & Cross-correlation\_PIV~ & $0.479$ & $0.341$ & $0.214$ \\
                          & RAFT\_PIV             & $0.339$ & $0.242$ & $0.152$ \\
SQG~                      & UnLiteFlowNet\_PIV    & $0.231$ & $0.184$ & $0.111$ \\
                          & Cross-correlation\_PIV~ & $1.059$ & $0.735$ & $0.388$ \\
                          & RAFT\_PIV             & $0.197$ & $0.145$ & $0.068$ \\
uniform~                  & UnLiteFlowNet\_PIV    & $1.183$ & $0.677$ & $0.134$ \\
                          & Cross-correlation\_PIV~ & $0.179$ & $0.089$ & $0.010$ \\
                          & RAFT\_PIV             & $1.828$ & $1.749$ & $0.362$ \\
\end{tblr}
\end{table}

\subsection{Event-based Imaging Velocimetry Application}

To evaluate the quality of the generated event data, we employed three representative methods: EBIV~\cite{willert2022event}, E-RAFT~\cite{GehrigEraft}, and Contrast-Maximization~\cite{Gallego_2018_CVPR}. For all three methods, we utilized the complete set of event data collected within a $15ms$ time window.

\begin{figure}[!hptb]
        \centering
    	 \includegraphics[width=\textwidth]{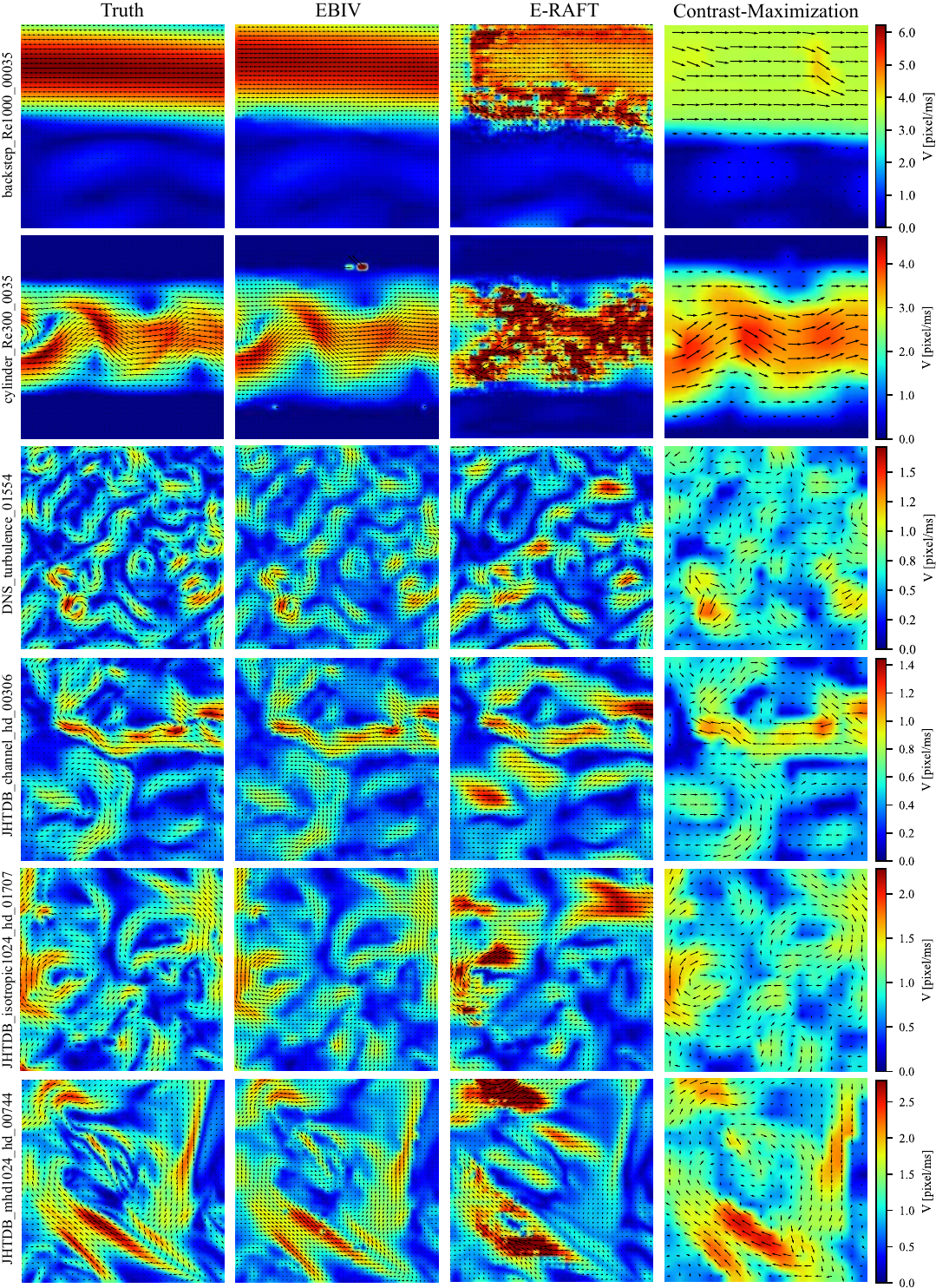}
	 \caption{The visualization of Event Application. 
	 }
        \label{fig:EventAssess}
\end{figure}

Figure~\ref{fig:EventAssess} shows the evaluation results of six different flow field datasets. Overall, the tested methods are able to reconstruct the basic characteristics of the flow fields. Among them, the EBIV method performs relatively well, but still exhibits distortions in event-sparse regions—for instance, outliers appear in the upper region of cylinder\_Re300\_00351. As for E-RAFT, due to the lack of algorithmic tuning in our experiments, its performance is suboptimal. Nevertheless, the results highlight the great potential of deep learning networks in processing event-based information. In contrast, the Contrast-Maximization method suffers from long computation times and limited accuracy. As shown in the figure, when the computational grid is coarse, it can only roughly capture the overall flow pattern and fails to estimate high-speed motion accurately, such as in backstep\_Re1000\_00035. Overall, We hope this dataset can serve as a strong foundation for developing faster and more accurate algorithms, particularly for event-driven deep learning approaches.


Similarly, the Table~\ref{Tab:EventMethod} summarizes the detailed statistical evaluation metrics of the three aforementioned methods. Overall, EBIV demonstrates higher computational accuracy than E-RAFT, and in some cases, its performance approaches that of PIV-based methods. Contrast-Maximization still has significant room for improvement: while its performance is relatively close to the other two methods in low-velocity flow fields, it falls short in high-speed scenarios. These evaluation metrics provide valuable benchmarks for developing event-based velocity estimation algorithms using our dataset. Furthermore, they point to a promising research direction—fusing image and event-based data to enhance computational accuracy through multimodal information integration.



\begin{table}
\centering
\caption{Quantitative comparison of different event methods evaluated on our proposed dataset. }
\label{Tab:EventMethod}
\begin{tblr}{
  colspec = {llccc},  
  cell{2}{1} = {r=3}{},
  cell{5}{1} = {r=3}{},
  cell{8}{1} = {r=3}{},
  cell{11}{1} = {r=3}{},
  cell{14}{1} = {r=3}{},
  cell{17}{1} = {r=3}{},
  cell{20}{1} = {r=3}{},
  cell{23}{1} = {r=3}{},
  cell{26}{1} = {r=3}{},
  hline{1,29} = {0.08em},  
  hline{2,5,8,11,14,17,20,23,26} = {-}{},
}
Flow type~                & Method                  & RMSE      & AEE       & AAE      \\
backstep~                 & EBIV                    & $0.249$   & $0.194$   & $0.114$  \\
                          & E-RAFT                  & $0.851$   & $0.747$   & $0.266$  \\
                          & Contrast-Maximization   & $1.236$   & $0.876$   & $0.216$  \\
cylinder~                 & EBIV                    & $0.591$   & $0.165$   & $0.647$  \\
                          & E-RAFT                  & $0.972$   & $0.765$   & $0.895$  \\
                          & Contrast-Maximization   & $0.484$   & $0.318$   & $0.699$  \\
DNS\_turbulence~          & EBIV                    & $0.632$   & $0.486$   & $0.301$  \\
                          & E-RAFT                  & $0.297$   & $0.249$   & $0.405$  \\
                          & Contrast-Maximization   & $0.854$   & $0.682$   & $0.422$  \\
JHTDB\_channel~           & EBIV                    & $0.203$   & $0.165$   & $0.019$  \\
                          & E-RAFT                  & $0.681$   & $0.591$   & $0.293$  \\
                          & Contrast-Maximization   & $0.758$   & $0.671$   & $0.067$  \\
JHTDB\_channel\_hd~       & EBIV                    & $0.122$   & $0.092$   & $0.183$  \\
                          & E-RAFT                  & $0.224$   & $0.192$   & $0.273$  \\
                          & Contrast-Maximization   & $0.303$   & $0.234$   & $0.365$  \\
JHTDB\_isotropic1024\_hd~ & EBIV                    & $0.290$   & $0.202$   & $0.150$  \\
                          & E-RAFT                  & $0.418$   & $0.301$   & $0.326$  \\
                          & Contrast-Maximization   & $0.513$   & $0.406$   & $0.307$  \\
JHTDB\_mhd1024\_hd~       & EBIV                    & $0.310$   & $0.199$   & $0.128$  \\
                          & E-RAFT                  & $0.507$   & $0.392$   & $0.373$  \\
                          & Contrast-Maximization   & $0.597$   & $0.448$   & $0.287$  \\
SQG~                      & EBIV                    & $0.795$   & $0.561$   & $0.307$  \\
                          & E-RAFT                  & $1.112$   & $0.997$   & $0.874$  \\
                          & Contrast-Maximization   & $0.905$   & $0.719$   & $0.344$  \\
uniform~                  & EBIV                    & $0.061$   & $0.060$   & $0.005$  \\
                          & E-RAFT                  & $0.048$   & $0.041$   & $0.003$  \\
                          & Contrast-Maximization   & $0.977$   & $0.910$   & $0.104$  
\end{tblr}
\end{table}

\section{Conclusion}


In this paper, we propose FED-PV, a bimodal data generation framework. By incorporating an extended computational domain and minimal iterative step design, FED-PV achieves high-fidelity modeling of particle motion under a steady flow field. Based on this framework, we construct an open-source dataset comprising $9$ typical flow field categories with a total size of $350GB$. Through systematic evaluations using multiple PIV and PEV algorithms, we provide key performance metrics of the dataset (RMSE, AEE and AAE) across a range of mainstream particle velocimetry methods. Given the current lack of research on multimodal particle velocimetry, the open-source framework and the accompanying dataset we provide (project repository: \url{https://github.com/DreamFuture6/FED-PV}) serve as a valuable resource to support future studies in high-performance multimodal particle velocimetry.

\begin{acknowledgments}
We would like to thank Jeremiah Hu and Wang Wei for providing support with the RAFT-PIV data used in this work. We would like to thank PhD candidate Jia Ai for the valuable discussions and helpful suggestions on the experimental design and methodology. We also express our gratitude to the ISPIV organizing committee. This work was supported by the Hubei Provincial Natural Science Foundation of China (Grant No. 2023AFB128).
\end{acknowledgments}
\FloatBarrier
\bibliography{MPV_References}

\end{document}